\documentclass{article}

\usepackage{arxiv}

\usepackage[utf8]{inputenc} 
\usepackage[T1]{fontenc}    
\usepackage{hyperref}       
\usepackage{url}            
\usepackage{booktabs}       
\usepackage{amsfonts}       
\usepackage{nicefrac}       
\usepackage{microtype}      
\usepackage{lipsum}	

\usepackage{graphicx}
\usepackage{amsmath,amssymb,enumerate,subfigure,textcase}
\usepackage{booktabs}
\usepackage{mathtools}
\usepackage{multirow}
\usepackage{array}
\usepackage{booktabs}
\usepackage{bigstrut}
\usepackage{float}
\usepackage{setspace}
\usepackage{balance}

\begin{document}

\title{Video Compression Coding via Colorization:\\A Generative Adversarial Network (GAN)-Based Approach}



\author{
  Zhaoqing Pan \\
  School of Computer and Software\\
  Nanjing University of Information Science and Technology\\
  Nanjing 210044, China \\
  \texttt{zhaoqingpan@nuist.edu.cn} \\
   \And
 Feng Yuan \\
  School of Computer and Software\\
  Nanjing University of Information Science and Technology\\
  Nanjing 210044, China \\
  \texttt{yuanfeng@nuist.edu.cn} \\
\And
  Jianjun Lei\thanks{corresponding author} \\
  School of Electrical and Information Engineering\\
  Tianjin University\\
  Tianjin 300072, China \\
   \texttt{Email: jjlei@tju.edu.cn}\\
   \And
   Sam Kwong\\
   Department of Computer Science\\
   City University of Hong Kong\\
   Hong Kong, China\\
   \texttt{cssamk@cityu.edu.hk}
}

\maketitle

\begin{abstract}
Under the limited storage, computing and network bandwidth resources, the video compression coding technology plays an important role for visual communication. To efficiently compress raw video data, a colorization-based video compression coding method is proposed in this paper. In the proposed encoder, only the video luminance components are encoded and transmitted. To restore the video chrominance information, a generative adversarial network (GAN) model is adopted in the proposed decoder. In order to make the GAN work efficiently for video colorization, the generator of the proposed GAN model adopts an optimized MultiResUNet, an attention module, and a mixed loss function. Experimental results show that when compared with the H.265/HEVC video compression coding standard using all-intra coding structure, the proposed video compression coding method achieves an average of 72.05\% BDBR reduction, and an average of 4.758 dB BDPSNR increase. Moreover, to our knowledge, this is the first work which compresses videos by using GAN-based colorization, and it provides a new way for addressing the video compression coding problems.
\end{abstract}


\section{Introduction}
With the development of imaging and display technologies, the high definition video becomes more and more popular in our life. However, with the increased number of resolution, frame rate, sampling precision, etc., the volume of raw video data increases significantly. The huge raw video data is a challenge for signal processing, storage, and transmitting. Hence, the efficient video compression coding technology becomes vital important for the video to be widely used in multimedia applications.

In the past three decades, video compression coding standards have been developed rapidly, which resulted in H.26X \cite{h265}, MPEG \cite{mpeg}, VC-1 \cite{vc1}, AVS \cite{avs}, and so on. They directly compress the luminance and chrominance components of raw video without any preprocessing, however, with the increased resolution, the video data required for compression also increases dramatically, which results in that the volume of compressed video is still large. Hence, the compression efficiency of current video compression coding standards should be further improved.

To efficiently compress raw video, we propose a new video compression coding framework in this paper, that the compression efficiency of traditional video coding standards is significantly improved by using the colorization-based methods, in which the encoder only compresses and transmits the video luminance components, and the chrominance information is recovered by a generative adversarial network (GAN)-based colorization method at the decoder. Overall, the contributions of this paper are summarized and listed as follows.
\begin{enumerate}
  \item[(1)] The problem of video chrominance component compression is converted to the grayscale video colorization problem.
  \item[(2)] To enhance the learning ability of the proposed GAN model, an optimized MultiResUNet is adopted as the backbone of the proposed GAN's generator.
  \item[(3)] To improve the generation capacity of the proposed GAN model, a mixed loss function is proposed for the generator.
  \item[(4)] To our knowledge, this is the first work which compresses videos by using GAN-based colorization, and it provides a new way for addressing the video compression coding problems. Particularly, when using a large quantization parameter (QP), the proposed video compression coding method can efficiently enhance the objective and subjective qualities of the compressed video, while the visual quality of video compressed by traditional video compression coding methods will degrade dramatically.
\end{enumerate}

The rest of this paper is organized as follows. In section \ref{sec2}, the related works are introduced. The details of proposed GAN-based video colorization for efficient video compression coding are presented in Section \ref{sec3}. The performance of proposed video compression coding method is given in Section \ref{sec4}. Finally, Section \ref{sec5} concludes this paper.

\section{Related Works}
\label{sec2}

GANs \cite{Gan} are one kind of probability generative model, which can extract the internal statistical rules of given observation data, and generate new data samples consistent with the distribution of observation data. Furthermore, GANs have the advantages of flexible structure and low complexity. In recent years, GANs have been widely used for multimedia processing, and the applications are summarized as that image super-resolution \cite{srgan}, image translation \cite{imageT}, video content analysis \cite{msgan}\cite{vgan}, natural language processing \cite{seqgan}\cite{ctpgan}, unmanned vehicle \cite{cargan}, medicine processing \cite{mgan}\cite{fbgan}, and so on. These GANs models have achieved excellent performance in their used applications, however, they are not suitable for directly using in video colorzation and compression, due to these GANs didn't consider the characteristics of video colorzation and compression problems.

Colorization is an old topic in computer vision, which aims to add colors to grayscale images, and make the colorized images seem meaningful. These image colorization methods can be roughly classified into three categories: scribble-based colorization, example-based colorization, and learning-based colorization. The scribble-based colorization methods \cite{c1}-\cite{c3} require user to provide large quantity manual labeling, then the hand-graffiti color spreads around to the entire image. Since the color information of the grayscale video cannot be randomly scribbled, these scribble-based colorization methods cannot be used in video compression coding system. The example-based colorization methods \cite{ebc2}-\cite{ebc4} extract color information from the reference images, and then transfer it to target grayscale images. These methods still have some limitations such as insufficient coloring accuracy, they still require the user to participate in the colorization process. The learning-based colorization methods \cite{lc1}-\cite{lc5} automatically learn the mapping relationship between raw grayscale images and corresponding color images from a large sample set. Even though these learning-based can automatically colorize the grayscale images, they are not suitable for to be used in video compression coding system due to the chrominance information must be accurately colorized in video coding process.

\begin{figure*}[htbp]
  \centering
  \includegraphics[width=5in]{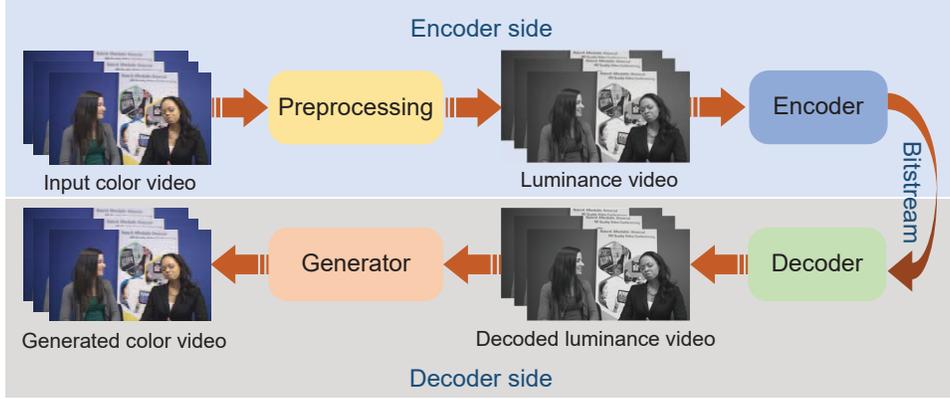}\\
  \caption{The working mechanism of the proposed video compressor.}\label{fig1}
\end{figure*}

\section{Proposed GAN-Based Video Colorization for Efficient Video Compression Coding}
\label{sec3}

\subsection{Motivation and Problem Modeling}
Since the human visual system is less sensitive to chrominance than to luminance, the YCbCr color space is an efficient method for representing color image in video compression coding systems. Y indicates the luminance component, Cb and Cr represent the chrominance components. There are three YCbCr sampling formats for Y, Cb, and Cr, including 4:2:0, 4:2:2, and 4:4:4, the numbers represent the sampling rate of each component in the horizontal direction. For example, the 4:2:2 sampling format indicates that for every 4 luminance samples in the horizontal direction have 2 Cb and 2 Cr samples. For compressing the video, its luminance and chrominance components are all compressed in traditional video compression coding systems, this compression coding process is formulated in Eq. (\ref{eq1}),
\begin{equation}
\left\{
  \begin{array}{lr}
    v=C(Y, Cb, Cr),\\
    v^{'}=D(v),
  \end{array}
\right.
\label{eq1}
\end{equation}
where $C()$ represents the video encoder; $Y$ is the video luminance signal, $Cb$ and $Cr$ are the video chrominance signals; $v$ means the compressed video; $D()$ indicates the video decoder; $v^{'}$ is the decoded video. Due to that these traditional video compressors encode all video signals, the volume of compressed video is still large when the resolution of raw video increases.

To improve video compression coding efficiency, we propose a new video compression coding mechanism that the video compressor only encodes and transmits the video luminance component, and the chrominance information is automatically generated in the decoder, the working mechanism of the proposed video compressor is shown in Fig. \ref{fig1}. In other words, the task of encoding chrominance components, as shown in Eq. (\ref{eq1}), is transferred into an image-to-image translation task, as shown in Eq. (\ref{eq2}),
\begin{equation}
\left\{
  \begin{array}{lr}
   v=C(Y),\\
   v^{'}=G(v),
  \end{array}
\right.
\label{eq2}
\end{equation}
where $C()$ is the video encoder; $Y$ is the video luminance signals; $v$ denotes the compressed luminance signals; $G()$ represents the color video generator; $v^{'}$ is the generated color video. Since the proposed compressor only encodes the luminance signals, the volume of compressed video will be decreased significantly. For example, the data volume of proposed encoder is 2/3 that of the traditional encoder for 4:2:0 color sampling pattern.

\begin{table}[htbp]
\centering
\caption{Feature size of each layer in the proposed GAN model}
\scalebox{0.9}{
\begin{tabular}{clcl}
\toprule
Layer & Feature Size&Layer&Feature Size \\ \midrule
P1, M1, A1&$W\times$$H\times C$ &D1& $W\big / 8\times H \big / 8\times 4C$ \\
P2, M2, A2&${W}\big/{2}\times{H}\big/{2}\times C$& D2&${W}\big/{4}\times{H}\big/{4}\times 4C$\\
P3, M3, A3&${W}\big/{4}\times{H}\big/{4}\times 2C$&D3&${W}\big/{2}\times{H}\big/{2}\times 2C$\\
P4, M4, A4&${W}\big/{8}\times{H}\big/{8}\times 2C$&D4&$W\times H\times 2C$\\
C1&${W}\big/{2}\times{H}\big/{2}\times C$&C2&${W}\big/{4}\times{H}\big/{4}\times C$\\
C3&${W}\big/{8}\times{H}\big/{8}\times C$&C4&${W}\big/{8}\times{H}\big/{8}\times C$\\
C5&${W}\big/{8}\times{H}\big/{8}$&&\\
  \bottomrule
\end{tabular}}
\label{size}
\end{table}

\begin{figure*}[htbp]
  \centering
  \includegraphics[width=6in]{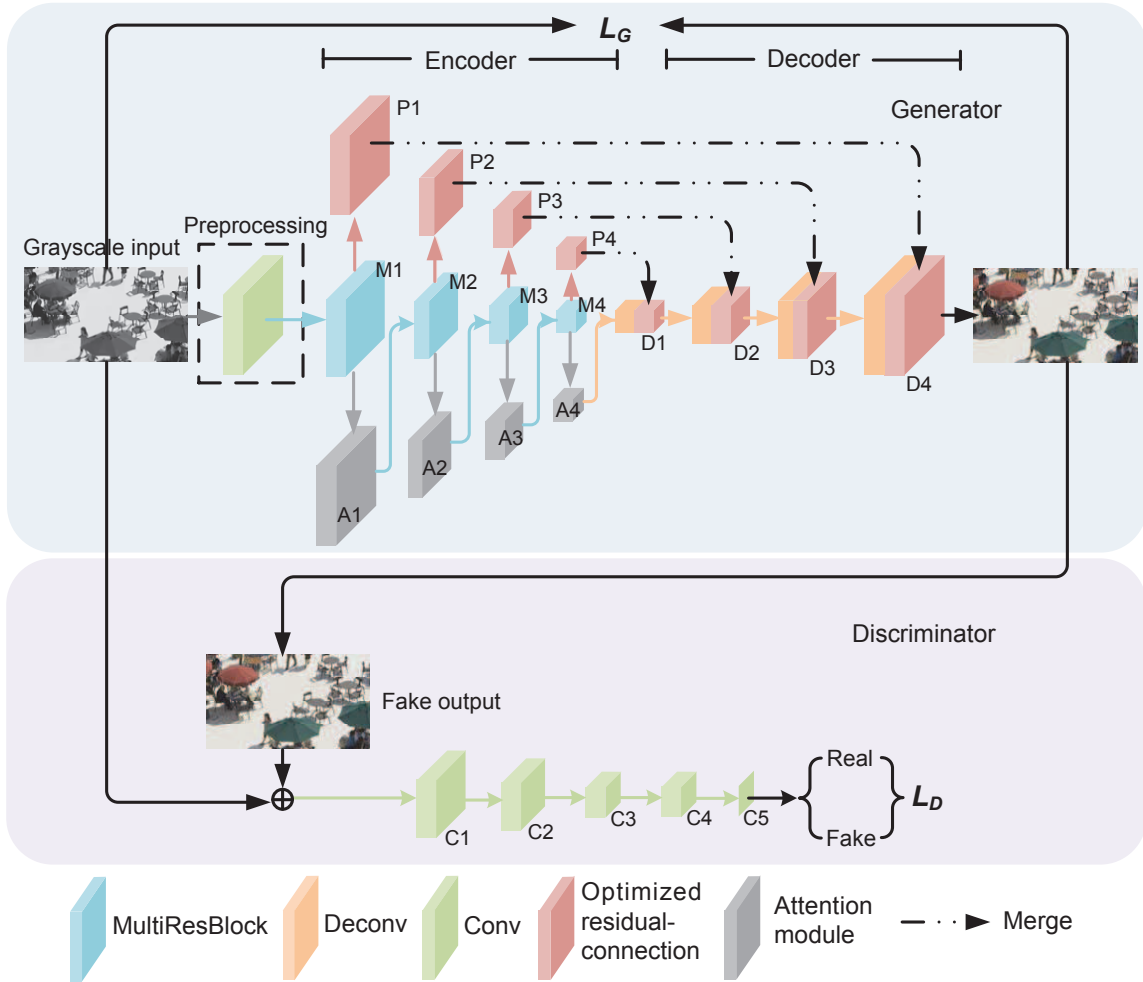}\\
  \caption{The architecture of the proposed GAN model.}\label{gan}
\end{figure*}

\subsection{Proposed GAN Model for Video Colorization}
GANs consist of two parts: generator and discriminator. The generator captures the data distribution, and generates fake data that has quite similar distribution with the real data. The discriminator aims to distinguish whether the generated sample is fake or comes from the real data. The regular GANs can generate real data, however, there is no any priori information used in the network, the generation process is liberal and uncontrollable. Moreover, one remarkable characteristic of video is that the video content is highly consistent in time domain. In order to make GANs well handle the video colorization task, a supervised GAN model is adopted for constraining the generation process, and colorizing the grayscale video accurately. The architecture of the proposed GAN model is shown in Fig. \ref{gan}. The feature size of each layer is listed in Table \ref{size}. In order to learn more global feature information, and enhance the learning ability of the proposed network, an optimized MultiResUNet is proposed, which is used as the backbone of the generator. In addition, an attention module \cite{sa} is applied into the optimized MultiResUNet for improving the colorization performance. Finally, a mixed loss function is proposed for further enhancing the generation capacity of the proposed GAN model. The details are discussed as follows.

\subsubsection{Optimized MultiResUNet}

\begin{figure}[htbp]
  \centering
  \includegraphics[width=3.5in]{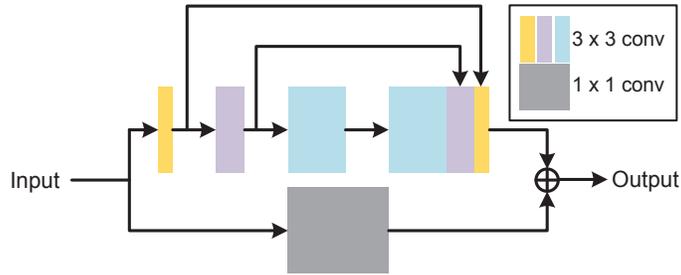}\\
  \caption{The architecture of the MultiResBlock.}\label{mb}
\end{figure}

MultiResUNet \cite{multiresunet} is an encoder-decoder model, and its skip-connection operation efficiently uses the feature information obtained by the encoder, which makes the deconvolution process more stable. As an improvement of the U-Net \cite{unet}, the MultiResUNet can efficiently handle the task of feature representation and transformation, and it has addressed the semantic gap problem between the corresponding levels of encoder-decoder in U-Net. The excellent performance of MultiResUNet benefits from the operation named multi-scale convolution operation. The multi-scale convolution operation is implemented by a MultiResBlock which connects the convolution layers in tandem. The architecture of the MultiResBlock is shown in Fig. \ref{mb}. The MultiResBlock contains a two-way 3$\times$3 convolution operation, and a three-way convolution operation. By these two operations, much more diverse features can be obtained to enhance the learning ability of the model.

\begin{figure}[htbp]
  \centering
  \includegraphics[width=3.5in]{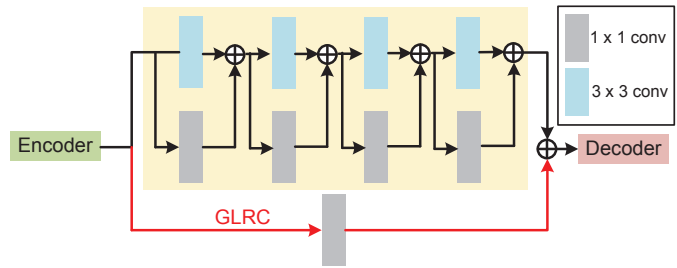}\\
  \caption{The architecture of the optimized residual connection.}\label{rc}
\end{figure}

In addition, a Residual Connection (RC) is used in the MultiResUNet, which makes the information obtained by the encoder become reasonable and effective. The regular RC consists of four residual blocks, and each residual block is made up of a 3$\times$3 convolution operation as well as a 1$\times$1 convolution shortcut. To further improve the learning performance of RC, we proposed an optimized RC, in which a Global Long Residual Connection (GLRC) is added, and the architecture of the optimized RC is presented in Fig. \ref{rc}. By jointly use of the global information and the features obtained by the residual blocks, the features which are transferred from the encoder to the decoder become more abundant and effective. The optimized RC is described as,
\begin{equation}
\left\{
  \begin{array}{lr}
   f_{o}=f_{GLRC}+R_{4},\\
   R_{4}=R_{1}\bigg( R_{1} \Big( R_{1}\big (R_{1}\big)\Big) \bigg),\\
   R_{1}=f_{3}+f_{1},
  \end{array}
\right.
\label{orse}
\end{equation}
where$ f_{o}$ is the final output features obtained by our optimized RC; $f_{GLRC}$ represents the features obtained by the proposed GLRC; $R_{4}$ stands for the four residual blocks, and $R_{1}$ means one residual block; $f_{1}$ and $f_{3}$ indicate the $1\times1$ and $3\times3$ convolution operations, respectively.

\subsubsection{Attention Module}

To strengthen the learning ability of the proposed network, and make the proposed network well capture the long-range dependence of features, the self-attention mechanism \cite{sa} is adopted. By using the self-attention mechanism, the generated color will appear continuously and consistently. The self-attention mechanism obtains its attention map through a series of matrix operations, then the attention map is used to operate with the original input feature map to generate the final transformed output feature map. The self-attention mechanism is formulated in Eq. (\ref{sae}),
\begin{equation}
\left\{
  \begin{array}{lr}
a= f(x)\bigotimes g(x), \\
o=softmax(a)\bigotimes h(x),\\
y=\lambda\cdot o+x,\\
  \end{array}
\right.
\label{sae}
\end{equation}
where $f$, $g$, and $h$ indicate the 1$\times$1 convolution operation, the attention mechanism applies three copies of the input feature map to perform the 1$\times$1 convolution operation; $a$ stands for the attention map obtained by the matrix operations with value map $f(x)$ and key map $g(x)$; the $softmax$ ensures the adopted attention information and the input feature map are in the same order of magnitude, and prevents the attention information becoming noise; $\lambda$ is the weight factor of the attention information; $y$ is the output feature map.

\subsubsection{Proposed Mixed Loss Function}

The loss function is used to evaluate the distortion between the data generated by the GAN model and the real data. A proper loss function can efficiently improve the learning ability of the network, and accelerate the learning convergence of the network. Video coding is a lossy data compression method, which will bring large distortion during the encoding process when a large QP is used, and a large distortion will significantly effect the colorization performance. To tackle this problem, we enhance the quality of the encoded video during the colorization process. The standard GAN loss only considers the high-frequency information in an image, while the low-frequency information is also important in human visual system. Based on this observation, a mixed loss function is proposed in this paper, and it consists of a GAN loss, a Mean Square Error (MSE) loss, a content loss \cite{contentloss}, as well as an optimized color loss. The details of each loss are introduced as follows.

\textbf{GAN Loss.} The GAN loss can ensure the generated image and the target image consistent in image level. Although the generated image has fake high-frequency information, the GAN loss can greatly improve the subjective quality of the generated image. It is defined in Eq. (\ref{ganloss}),
\begin{equation}
L_{GAN}=-\log D\big(G(Y)\big),
\label{ganloss}
\end{equation}
where $Y$ denotes the grayscale input; $D$ and $G$ represent the discriminative model and the generative model, respectively.

\textbf{MSE Loss.} In order to improve the objective quality of the generated image, the quality of the generated image should be measured at pixel level. Therefore, the MSE loss is used, and it is formulated in Eq. (\ref{mseloss}),
\begin{equation}
L_{MSE}=\|G(Y)-X\|_{l_{2}},
\label{mseloss}
\end{equation}
where $X$ represents the target color image. The MSE loss can efficiently improve the smoothness of the generated image. Combined with the GAN loss, the MSE loss can correct the high-frequency information, and effectively guarantee the subjective quality of the generated image.

\textbf{Content Loss.} The MSE loss can improve the objective quality of the generated image, while the generated image looks blurry. To tackle this problem, the content loss \cite{contentloss} is adopted in our proposed method, which makes use of the pre-trained VGG-19 network. The target image and generated image are input into the pre-trained VGG-19 network, as a result, the perceptual quality of generated image is greatly improved by evaluating their feature consistency. The content loss is defined in Eq. (\ref{contentloss}),
\begin{equation}
L_{content}=\frac{1}{C_{j}H_{j}W_{j}}\|\psi_{j}\big(G(Y)\big)-\psi_{j}(X)\|_{l_{1}},
\label{contentloss}
\end{equation}
where $\psi_{j}$ indicates the $j_{th}$ layer in pre-trained VGG-19 network; $C_j$, $H_j$ and $W_j$ represent the number, height and width of the feature map, respectively.

\textbf{Optimized Color Loss.} To make the generated image be consistent with the color distribution of its target image, a Gaussian blur is utilized, and the Euclidean distance between Gaussian filtering representations is computed \cite{colorloss}.  The Gaussian blur can remove the high-frequency information, and it measures the differences between two images at the fuzzy color level rather than pixel level. The Gaussian filter is represented as,
\begin{equation}
G(k,l)=\theta\cdot\exp \Big(-\frac{{(k-\mu_{x})}^{2}}{2\sigma_{x}} -\frac{{(l-\mu_{y})}^{2}}{2\sigma_{y}}\Big),
\end{equation}
where $\mu_x$ and $\mu_y$ are set to 0, $\sigma_x$ and $\sigma_y$ are defined to 3. To address the problem of brightness reduction caused by MSE loss, the parameter $\theta$ is optimized in this paper. Then, the color loss is defined as,
\begin{equation}
L_{color}={\parallel g(Y)\ast G_{o}-X\ast G_{t}\parallel}^{2}_{2},
\end{equation}
where $\ast$ represents the filtering operation; $g(Y)$ and $X$ indicate the generate image and its target image, respectively; $G_{o}$ and $G_{t}$ are Gaussian filters with different $\theta$, and the values of $\theta$ for $G_{o}$ and $G_{t}$ are set to 0.062 and 0.065, respectively, in this paper.

Finally, in order to achieve the optimal subjective and objective quality, the proposed mixed loss function is defined as
\begin{equation}
L_{f}=\alpha_{1}\cdot L_{GAN}+\alpha_{2}\cdot L_{MSE}+\alpha_{3}\cdot L_{content}+\alpha_{4}\cdot L_{color},
\end{equation}
where the content loss is based on the features obtained by the relu\_5\_4 layer of the VGG-19 network; $\alpha_{1}$, $\alpha_{2}$, $\alpha_{3}$, and $\alpha_{4}$ are four weight factors for the GAN loss, MSE loss, content loss, and optimized color loss, respectively. The philosophy of how to determine the optimal weight factor for each loss is that the loss of each term should be limited to the same order of magnitude. Based on experiments, when $\alpha_{1}$, $\alpha_{2}$, $\alpha_{3}$, and $\alpha_{4}$ are set to 1, 100, 1000, and 100, respectively, the proposed network achieves the best performance.

\section{Experimental Results}
\label{sec4}

\subsection{Experimental Setups}
We select the HEVC reference software HM 16.0 \cite{HM16} as the software platform of our proposed video compression coding method. The 4:0:0 format video stream and the trained model are simultaneously sent to the decoder after encoding process is finished, then the decoder will decode the stream, and colorize the 4:0:0 format video based on the trained model. The detailed experimental setups are given as follows.

\subsubsection{Data Preparation}
In this paper, we select 8 conferencing video, including six sequences with resolution 1280$\times$720 (Fourpeople, Johny, KristenAndSara, vidyo1, vidyo3, and vidyo4), and two sequences with resolution 352$\times$288 (Silent, and Mother-daughter). During the encoding process, all frames are classified into multiple group of pictures (GOP), and the GOP size equals to 6. The first frame of each GOP is encoded by the original HM 16.0 with 4:2:0 color samping under the all-intra coding structure, and these frames are also used for training. The remaining frames of each GOP are encoded by the proposed compression method.

\subsubsection{Training}
The learning rate is set to 0.0002, and the Adam optimization method is used to minimize the loss function. The training platform is windows 7 operating system with NVIDIA 1080Ti GPU.

\begin{table*}[htbp]
\centering
\caption{Encoding results}
\scalebox{0.88}{
\begin{tabular}{lccccccccc}
\toprule
\multirow{2}{*}{Sequence} & \multirow{2}{*}{QP} & \multicolumn{ 2}{c}{HM 16.0} & \multicolumn{ 2}{c}{Proposed} & \multicolumn{ 1}{c}{$\Delta$BR} & \multicolumn{ 1}{c}{$\Delta$PSNR} & \multicolumn{ 1}{c}{BDBR} & \multicolumn{ 1}{c}{BDPSNR} \\ \cmidrule(r){3-4} \cmidrule(r){5-6}
\multicolumn{ 1}{l}{} & \multicolumn{ 1}{c}{} & Bitrate (kbps) & PSNR (dB) & Bitrate (kbps) & PSNR (dB) & \multicolumn{ 1}{c}{(\%)} & \multicolumn{ 1}{c}{(dB)} & \multicolumn{ 1}{c}{(\%)} & \multicolumn{ 1}{c}{(dB)} \\ \midrule
\multirow{4}{*}{Silent} & 27 & 5018.75 & 35.232 & 4490.90 & 39.849 & -10.52 & 4.618 & \multirow{4}{*}{-90.46} & \multirow{4}{*}{6.811} \\
& 32 & 2605.17 & 32.350 & 2311.29 & 38.297 & -11.28 & 5.948 & &  \\
& 37 & 1279.90 & 29.888 & 1110.60 & 36.913 & -13.23& 7.025 &  &  \\
 & 42 & 632.48 & 27.751 & 533.18 & 35.543 & -15.70 & 7.792 &  &  \\ \midrule
\multirow{4}{*}{Mother-daughter} & 27 & 2130.04 & 38.551 & 1942.74 & 40.371 & -8.79 & 1.820 & \multirow{4}{*}{-71.21} & \multirow{4}{*}{4.178} \\
 & 32 & 1179.45 & 35.781 & 1076.89 & 39.059 & -8.70 & 3.278 & &  \\
 & 37 & 625.42 & 33.238 & 561.30 & 37.710 & -10.25 & 4.473 &  &  \\
 & 42 & 321.11 & 30.872 & 281.46 & 36.247 & -12.35 & 5.375 &  & \\ \midrule
\multirow{4}{*}{Fourpeople} & 27 & 8445.52 & 37.441 & 7737.71 & 39.431 & -8.38 & 1.990 & \multirow{4}{*}{-94.95} & \multirow{4}{*}{5.044} \\
 & 32 & 5240.35 & 34.378 & 4831.90 & 38.321 & -7.79 & 3.943 & &  \\
 & 37 & 3039.62 & 31.434 & 2787.45 & 36.742 & -8.30 & 5.307 & & \\
& 42 & 1593.58 & 28.477 & 1441.66 & 35.132 & -9.53 & 6.656 &  &  \\ \midrule
\multirow{4}{*}{Johnny} & 27 & 4574.14 & 38.819 & 4052.26 & 41.697 & -11.41 & 2.877 & \multirow{4}{*}{-79.79} & \multirow{4}{*}{5.739} \\
& 32 & 2721.10 & 36.095 & 2407.06 & 40.675 & -11.54 & 4.579 & &  \\
 & 37 & 1597.91 & 33.561 & 1383.51 & 39.615 & -13.42 & 6.054 &  & \\
 & 42 & 939.50 & 30.946 & 799.05 & 38.204 & -14.95 & 7.258 & &  \\ \midrule
\multirow{4}{*}{KristenAndSara} & 27 & 5926.84 & 38.503 & 5328.15 & 39.893 & -10.10 & 1.390 & \multirow{4}{*}{-60.14} & \multirow{4}{*}{4.253} \\
 & 32 & 3686.68 & 35.640 & 3332.31 & 38.652 & -9.61 & 3.012 &  & \\
 & 37 & 2211.07 & 32.824 & 1976.24 & 37.264 & -10.62 & 4.439 & &  \\
& 42 & 1255.51 & 29.970 & 1101.93 & 35.953 & -12.23 & 5.983 & &  \\ \midrule
\multirow{4}{*}{Vidyo1} & 27 & 5761.00 & 38.425 & 5281.38 & 40.486 & -8.33 & 2.060 & \multirow{4}{*}{-65.56} & \multirow{4}{*}{4.557} \\
& 32 & 3525.47 & 35.627 & 3252.31 & 39.197 & -7.75 & 3.570 &  &  \\
& 37 & 2071.47 & 32.881 & 1905.44 & 37.675 & -8.02 & 4.793 &  & \\
 & 42 & 1129.64 & 30.066 & 1026.00 & 35.737 & -9.17 & 5.671 &  & \\ \midrule
\multirow{4}{*}{Vidyo3} & 27 & 6222.16 & 38.413 & 5911.37 & 40.266 & -4.99 & 1.853 & \multirow{4}{*}{-65.80} & \multirow{4}{*}{4.586} \\
& 32 & 3769.25 & 35.455 & 3610.25 & 39.262 & -4.22 & 3.807 &  &  \\
& 37 & 2067.38 & 32.459 & 1984.91 & 37.432 & -3.99 & 4.974 & &  \\
 & 42 & 952.38 & 29.493 & 907.66 & 35.022 & -4.70 & 5.530 &  &  \\ \midrule
\multirow{4}{*}{vidyo4} & 27 & 5699.43 & 38.373 & 5252.65 & 38.867 & -7.84 & 0.494 & \multirow{4}{*}{-48.48} & \multirow{4}{*}{2.897} \\
& 32 & 3340.29 & 35.537 & 3077.13 & 37.607 & -7.88 & 2.071 & & \\
& 37 & 1926.97 & 32.889 & 1757.99 & 35.952 & -8.77 & 3.063 &  & \\
& 42 & 1054.13 & 30.210 & 949.68 & 34.470 & -9.91 & 4.261 &  &  \\ \midrule
\multicolumn{2}{c}{Average}   & 2891.99 & 33.799 & 2637.64 & 38.048 & -9.51 & 4.249 & -72.05 & 4.758 \\ \bottomrule
\end{tabular}}
\label{ep1}
\end{table*}

\subsection{Encoding performance}
The encoding parameters of the original HM 16.0 are set as follows: (1) the all-intra coding structure is used; (2) four QPs are adopted, including 27, 32, 37, and 42; (3) the coding tree unit and quadtree depth are set to 64$\times$64 and 4, respectively; (4) the rate distortion optimization is enabled; (5) the default settings are used for the other coding parameters. The encoding results are summarized and tabulated in Table \ref{ep1}. In the Table, the $\Delta$BR and $\Delta$PSNR are computed as
\begin{equation}
\left\{
  \begin{array}{lr}
   \Delta BR=\frac{BR_{p}-BR_{o}}{BR_{o}}\times100\%,\\
   \Delta PSNR=PSNR_{p}-PSNR_{o},
  \end{array}
\right.
\label{brpsnr}
\end{equation}
where $BR_{p}$ and $PSNR_{p}$ indicate the bitrate and PSNR of our proposed compression method, respectively; $BR_{o}$ and $PSNR_{o}$ represent the bitrate and PSNR of the original HM16.0, respectively. The BDBR and BDPSNR are calculated according to \cite{bd}.

\begin{figure*}[htbp]
  \centering
  \includegraphics[width=5.8in]{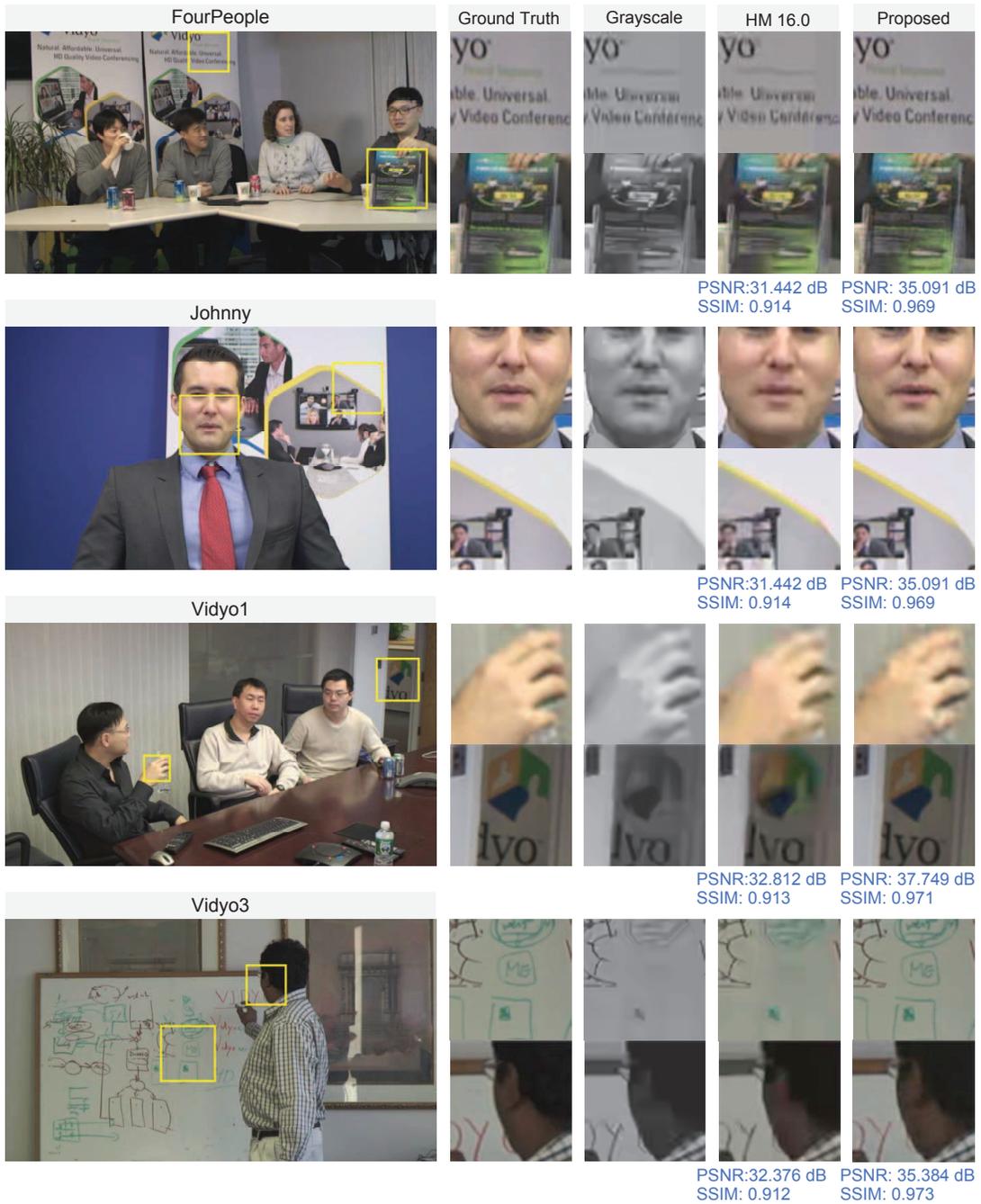}\\
  \caption{Video subjective quality comparison. }\label{subjective}
\end{figure*}

From Table \ref{ep1}, it can be seen that compared with the original HM16.0, the bitrate of the proposed method decreases from -3.99\% to -15.70\%, -9.51\% on average; and the PSNR of the proposed method increases from 0.494 dB to 7.325 dB, 4.249 dB on average. Meanwhile, the BDBR between the proposed method and original HM16.0 is from -48.48\% to -90.46\%, -72.05\% on average; and the BDPSNR between the proposed method and original HM16.0 is from 2.897 dB to 6.811 dB, 4.758 dB on average. From these values, we can observed that compared with the traditional HEVC encoder, the proposed video compression coding method achieves significantly better compression efficiency. Especially, the proposed compression coding method can efficiently address the colorization task for grayscale videos with large QPs. These values also prove that the proposed GAN model works efficiently for video colorization in video compression coding process.

\subsection{Comparisons on Subjective Visual Quality}
To intuitively present the performance of the proposed colorization-based video compression coding method, the subjective visual quality is compared, and shown in Fig. \ref{subjective}. In Fig. \ref{subjective}, the results of the 38th frame of each video sequence (FourPeople, Johnny, Vidyo1, and Vidyo3) are presented. The ``ground truth'' presents the uncompressed video, the ``grayscale'' shows the encoded grayscale video with QP37, the ''HM 16.0'' means the color video encoded by HM 16.0 with QP37, and the ``Proposed'' indicates the video compressed by the proposed colorization-based video compression coding method with QP37. We can see that the video encoded by HM 16.0 loses much details when using a large QP. By contrary, the proposed colorization-based video compression coding method can well retain the video content information when a large QP is used. Moreover, the values of PSNR and SSIM of the proposed method are greatly better than that of the traditional HEVC encoder. From these figures, it can be concluded that the proposed video compression coding method efficiently improves the subjective quality of the compressed video, as compared with the traditional HEVC encoder.

\subsection{Ablation Study}

\subsubsection{Ablation Study for the Proposed Residual Connection}
To further optimize the learning performance of RC in the MultiResUNet, an optimized RC is proposed by adding a GLRC. To prove the efficiency of the optimized RC, the performance in terms of PSNR and SSIM of the proposed RC is compared with the original RC, and the experimental results are presented in Table \ref{prp}.

\begin{table}[htbp]
\centering
\caption{Ablation Study results of the Residual Connection (QP=37)}
\scalebox{0.95}{
\begin{tabular}{lcccc}
\toprule
\multirow{2}{*}{Sequence} & \multicolumn{ 2}{c}{Original RC} & \multicolumn{ 2}{c}{Proposed RC} \\ \cmidrule(r){2-3} \cmidrule(r){4-5}
 & PSNR (dB) & SSIM & PSNR (dB) & SSIM \\ \midrule
Fourpeople & 36.640 & 0.970 & 36.742 & 0.970 \\
Johnny & 39.395 & 0.958 & 39.615 & 0.963 \\
Vidyo1 & 36.953 & 0.969 & 37.675 & 0.971 \\
Vidyo3 & 37.390 & 0.977 & 37.432& 0.977 \\ \midrule
Average & 37.594 & 0.969 & \textbf{37.866} & \textbf{0.970} \\ \bottomrule
\end{tabular}}
\label{prp}
\end{table}

It can be observed that the performance of the proposed network is significantly improved by using the optimized RC, the PSNR of the proposed RC with GLRC increases an average of 0.272 dB, and the SSIM of the proposed RC has an average of 0.001 increase. From these results, we conclude that the the proposed GLRC plays an important role in improving the learning ability of the proposed network.

\subsubsection{Ablation Study for the Attention Module}

The self-attention (SA) mechanism is used in our proposed network to strengthen the network's learning ability. To demonstrate the efficiency of the attention module, the network's performance of using and without using the SA is compared, and the comparison results are listed in Table \ref{sar}.

\begin{table}[htbp]
\centering
\caption{Ablation Study results of the self-attention mechanism (QP=37)}
\scalebox{0.95}{
\begin{tabular}{lcccc}
\hline
\multirow{2}{*}{Sequence} & \multicolumn{ 2}{c}{Network removing SA} & \multicolumn{ 2}{c}{Network using SA} \\ \cmidrule(r){2-3} \cmidrule(r){4-5}
 & PSNR (dB) & SSIM & PSNR (dB) & SSIM \\ \midrule
Fourpeople & 36.730 & 0.970 & 36.742 & 0.970 \\
Johnny & 39.129 & 0.960 & 39.615 & 0.963 \\
Vidyo1 & 37.380 & 0.972 & 37.675 & 0.971 \\
Vidyo3 & 37.018 & 0.975 & 37.432 & 0.977 \\ \midrule
Average & 36.730 & 0.970 & \textbf{37.866} & \textbf{0.970} \\ \bottomrule
\end{tabular}}
\label{sar}
\end{table}

It is seen from Table \ref{sar} that when the SA is used, the network's learning ability is significantly improved. The PSNR and SSIM of the proposed network without using the SA are 36.730 dB, and 0.970, respectively. The PSNR and SSIM of the proposed network using the SA are 37.866 dB, and 0.970, respectively. The PSNR value of the proposed network with using SA increases 1.136 dB, as compared with the proposed network removing the SA. From these values, it is concluded that the SA works efficiently for improving the learning ability of the proposed network.

\subsubsection{Ablation Study for the Proposed Mixed Loss Function}

The proposed mixed loss function consists of GAN loss, MSE loss, content loss, and optimized color loss. To testify the efficiency of the proposed mixed loss function, a group of comparative experiments have been performed. The detailed group information is shown in Table \ref{group}, and the experimental results are listed in Table \ref{as}.

\begin{table}[htbp]
\centering
\caption{Group for the loss functions}
\begin{tabular}{cc}
\toprule
Group & {Loss} \\ \midrule
G1 & $L_{GAN}$+100$\cdot$$L_{MSE}$ \\
G2 &$ L_{GAN}$+100$\cdot$$L_{MSE}$+100$\cdot$$L_{color}$ \\
G3 & $L_{GAN}$+100$\cdot$$L_{MSE}$+1000$\cdot$$L_{content}$ \\
G4 & $L_{GAN}$+100$\cdot$$L_{MSE}$+100$\cdot$$L_{color}$+1000$\cdot$$L_{content}$ \\ \bottomrule
\end{tabular}
\label{group}
\end{table}

From Table \ref{as}, it can be observed that the average PSNR and SSIM of G1 are 36.245 dB and 0.967, respectively. For G2, its average PSNR and SSIM are 37.218 dB and 0.969, respectively. The average PSNR and SSIM of G3 are 37.317 dB and 0.967, respectively. By using content loss and color loss, the performance of G2 and G3 is greatly better than that of G1. Finally, the loss function G4 achieves the best performance, its PSNR reaches up to an average of 37.866 dB, and its SSIM achieves an average of 0.970. These values reflect that the proposed loss function efficiently improves network performance.

\begin{table*}[htbp]
\centering
\caption{The performance of different loss functions (QP=37)}
\scalebox{0.9}{
\begin{tabular}{lcccccccc}
\toprule
\multirow{2}{*}{Sequence} & \multicolumn{ 2}{c}{G1} & \multicolumn{ 2}{c}{G2 } & \multicolumn{ 2}{c}{G3} & \multicolumn{ 2}{c}{G4} \\ \cmidrule(r){2-3}\cmidrule(r){4-5}\cmidrule(r){6-7}\cmidrule(r){8-9}
 & PSNR (dB) & SSIM & PSNR (dB) & SSIM & PSNR (dB) & SSIM & PSNR (dB) & SSIM \\ \midrule
Fourpeople & 36.379 & 0.968 & 36.546 & 0.970 & 36.481 & 0.969 & 36.742 & 0.970 \\
Johnny & 38.632 & 0.958 & 38.730 & 0.961 & 39.382 & 0.962 & 39.615 & 0.963 \\
Vidyo1 & 34.779 & 0.968 & 36.640 & 0.970 & 36.053 & 0.961 & 37.675 & 0.971 \\
Vidyo3 & 35.188 & 0.974 & 36.956 & 0.976 & 37.354 & 0.976 & 37.432 & 0.977 \\\midrule
Average&36.245&	0.967&37.218&0.969&37.317&	0.967&\textbf{37.866}&\textbf{0.970}\\ \bottomrule
\end{tabular}}
\label{as}
\end{table*}

\subsection{Comparison with the State-of-the-Art GANs}

To prove the outstanding performance of the proposed network in handling colorization problem, the proposed method is compared with 4 state-of-the-art GAN models, including Isola \cite{imageT}, Wang \cite{HDpix}, Zhu \cite{cycle}, and Isola \cite{imageT}+Zhang \cite{sa}. ``Isola \cite{imageT}+Zhang \cite{sa}'' represents the self-attention mechanism in Zhang \cite {sa} is used for Isola \cite{imageT}. Among these methods, Isola \cite{imageT} and Wang \cite{HDpix} are paired training methods, while Zhu \cite{cycle} and Isola \cite{imageT}+Zhang \cite{sa} are unpaired training methods. In these experiments, the grayscale videos are encoded by HM 16.0 with QP27. Since the input size of  Isola \cite{imageT} and Zhu \cite{cycle} should be square, the videos are cropped into square videos in this experiment. The PSNR and SSIM are used to evaluate the colorization performance of each method, and the experimental results are listed in Table \ref{cs}. We can see that the average PSNR and average SSIM of Isola \cite{imageT} are 34.066 dB, and 0.953, respectively. The Wang \cite{HDpix} achieves an average of 30.980 dB PSNR, and an average of 0.916 SSIM. For the Zhu \cite{cycle}, the PSNR and SSIM are 35.996 dB and 0.975, respectively. The average PSNR and average SSIM of the Isola \cite{imageT}+Zhang \cite{sa} are 35.518 dB and 0.959, respectively. For the proposed method, its average PSNR is 41.192 dB, and its average SSIM is 0.985. From these results, we observe that the proposed GAN model efficiently handling the video colorization for video compression coding.

\begin{table*}[htbp]
\centering
\caption{Comparisons on the State-of-the-Art GANs}
\scalebox{0.85}{
\begin{tabular}{lcccccccccc}
\toprule
\multirow{2}{*}{Sequence} & \multicolumn{ 2}{c}{Isola \cite{imageT}} & \multicolumn{ 2}{c}{Wang \cite{HDpix}} & \multicolumn{ 2}{c}{Zhu \cite{cycle}} & \multicolumn{ 2}{c}{Isola \cite{imageT}+Zhang \cite{sa}} & \multicolumn{ 2}{c}{Proposed} \\ \cmidrule(r){2-3}\cmidrule(r){4-5}\cmidrule(r){6-7}\cmidrule(r){8-9}\cmidrule(r){10-11}
 & PSNR (dB) & SSIM & PSNR (dB) & SSIM & PSNR (dB) & SSIM & PSNR (dB) & SSIM & PSNR (dB) & SSIM \\ \midrule
Silent & 32.553 & 0.920 & 29.860 & 0.886 & 38.489 & 0.979 & 32.499 & 0.926 & 38.976 & 0.987 \\
Mother-daughter & 30.600 & 0.922 & 33.237 & 0.904 & 33.685 & 0.961 & 33.234 & 0.929 & 40.671 & 0.986 \\
Fourpeople & 35.967 & 0.969 & 30.146 & 0.913 & 31.731 & 0.972 & 37.224 & 0.969 & 40.329 & 0.983 \\
Johnny & 37.252 & 0.962 & 29.924 & 0.900 & 37.616 & 0.975 & 38.174 & 0.971 & 43.418 & 0.994 \\
KristenAndSara & 34.792 & 0.965 & 30.469 & 0.922 & 37.104 & 0.981 & 36.311 & 0.971 & 40.699 & 0.982 \\
vidyo1 & 32.367 & 0.954 & 31.267 & 0.929 & 35.166 & 0.976 & 35.934 & 0.970 & 41.100 & 0.983 \\
vidyo3 & 34.527 & 0.972 & 31.725 & 0.949 & 34.073 & 0.975 & 36.229 & 0.974 & 41.913 & 0.985 \\
vidyo4 & 34.473 & 0.961 & 31.212 & 0.922 & 40.101 & 0.984 & 34.538 & 0.959 & 42.434 & 0.978 \\ \midrule
Average & 34.066 & 0.953 & 30.980 & 0.916 & 35.996 & 0.975 & 35.518 & 0.959 & \textbf{41.192} & \textbf{0.985} \\ \bottomrule
\end{tabular}}
\label{cs}
\end{table*}

\section{Conclusion}
\label{sec5}
To efficiently compress color video, we proposed a colorization-based video compression coding method, which transfer the traditional chrominance signals encoding problem to the grayscale video colorization problem. The proposed compressor only encodes and transmits the video luminance signals. To restore the video chrominance signals, a GAN model is adopted in the decoder, and the proposed GAN model uses an optimized MultiResUNet, an attention module, and a mixed loss function. Experimental results show that the proposed method achieves an average of 72.05\% BDBR reduction, and an average of 4.758 dB BDPSNR increase, as compared with traditional HEVC encoder. Moreover, the proposed method provides a new way for addressing the video compression coding problems.

\balance

\end{document}